\documentstyle[12pt]{article}
\newcommand{\be}{\begin{equation}}
\newcommand{\ee}{\end{equation}}
\newcommand{\eq}{\begin{equation}}
\newcommand{\en}{\end{equation}}
\newcommand{\bc}{\begin{center}}
\newcommand{\ec}{\end{center}}
\newcommand{\lsim}{\raisebox{0.3mm}{\em $\, <$} 
\hspace{-3.3mm} \raisebox{-1.8mm}{\em $\sim \,$}}
\newcommand{\gsim}{\raisebox{0.3mm}{\em $\, >$} 
\hspace{-3.3mm} \raisebox{-1.8mm}{\em $\sim \,$}}
\rightline{KEK-TH-396}
\rightline{KEK Preprint 94-17}
\rightline{TMUP-HEL-9402} 
\rightline{hep-ph 9405239}
\rightline{April 1994}
\begin{document} 
\vglue 0.5cm	
\bc
\Large\bf Testing the  Principle of Equivalence 
by Solar Neutrinos\\
\vglue 0.5cm 
\ec
\vglue 1cm	
\centerline{
Hisakazu Minakata$^{1,3}$ and  Hiroshi Nunokawa$^{2,3}$}
\vglue 0.5cm	
\centerline{$^1$\it Department of Physics, Tokyo Metropolitan 
University}
\centerline{\it 1-1 Minami-Osawa Hachioji, Tokyo 192-03 Japan}
\vglue 0.5cm	
\centerline{$^2$\it Theory Group, 
National Laboratory for High Energy Physics (KEK)}
\centerline{\it 1-1 Oho, Tsukuba-shi, Ibaraki-ken, 305 Japan}
\vglue 0.5cm	
\centerline{$^3$\it Institute for Nuclear Theory, University of 
Washington}
\centerline{\it Seattle, WA98195, USA}
\vglue 1.5cm	
\centerline{ABSTRACT}
\vglue 1cm	
We discuss the possibility of testing the principle of 
equivalence 
with solar neutrinos. If there exists a violation of the 
equivalence principle quarks and leptons with different flavors 
may not universally couple with gravity. The method we discuss 
employs a quantum mechanical phenomenon of neutrino oscillation 
to probe into the non-universality of the gravitational 
couplings of neutrinos. 
We develop an appropriate formalism to deal with neutrino 
propagation under the weak gravitational fields of the sun in the 
presence of the flavor mixing. We point out that solar neutrino 
observation by the next generation water Cherenkov detectors can 
improve the existing bound on violation of the equivalence 
principle by 3-4 orders of magnitude if the nonadiabatic 
Mikheyev-Smirnov-Wolfenstein mechanism is the solution to 
the solar neutrino problem.
\hfil\break
\newpage
\noindent
Experimental test of the principle of equivalence, one of the
fundamental building block of Einstein's general theory of 
relativity, now has history of more than a century. 
It was E\"otv\"os \cite{EPF} who carried out the celebrated 
torsion 
balance experiment that elevated the experimental test of the 
equivalence principle to the realm of precise measurements. 
E\"otv\"os and his collaborators obtained the bound 
$\eta\lsim3\times10^{-9}$ for the violation of the equivalence of 
inertial and gravitational masses.
Forty years later the E\"otv\"os-type torsion balance experiment 
was 
substantially improved by Dicke and his collaborators \cite{RKD}.
They noticed that sun's gravitational field affects this type of 
experiment by producing torque of 24 hours period in the presence 
of violation of the equivalence principle. With their extensive 
efforts in removing systematic uncertainties they achieved the 
accuracy $\eta\lsim3\times10^{-11}$, an impressive improvement 
over two orders of magnitude over E\"otv\"os'. Using the similar 
apparatus with Dicke et al.'s Braginski and Panov \cite{BP} 
reported the bound $\eta\lsim0.9\times10^{-12}$, the most 
stringent one to date.

In this paper we point out that by using the sun as the 
source of gravity {\it and at the same time} as the source of 
neutrino beam one can obtain much more stringent constraint 
for violation of the equivalence principle. 
We show that the optimal sensitivity one would achieve by 
the solar neutrino observation by the next-generation water 
Cherenkov detectors is $|\Delta f|\simeq10^{-15}-10^{-16}$, 
an improvement over 3-4 orders of magnitude than 
the E\"otv\"os-Dicke-type experiment.
Here $\Delta f$ is a measure for violation of the equivalence 
principle for neutrinos to be defined later.

A mild bound $|\Delta f| \lsim 10^{-3}$ for violation of the 
equivalence principle for neutrinos have been derived \cite{LKT} 
by using the small difference between arrival times of photons and 
neutrinos from SN1987A. The best bound deduced so far for 
microscopic objects is from the neutron free fall refractometry 
experiments which led to $|\eta| < 3 \times 10^{-4}$ \cite{neutron}.

Our discussion in this paper heavily relies on the basic observation 
by Gasperini \cite{GAS}. If Einstein's equivalence principle is 
violated gravity may not universally couple with neutrinos with 
different flavors. He pointed out that if this occurs neutrino 
oscillation similar to that of flavor mixing takes place and the 
effect can be detectable by experiments. After his observation 
several attempts have been made to sharpen his proposal and to 
examine the attainable sensitivity for violation of the equivalence 
principle.

Halprin and Leung \cite{HL} observed that one can use solar 
neutrinos to perform a sensitive test of the equivalence principle 
and they gave an order-of-magnitude estimation of the sensitivity, 
$|\Delta f|\sim10^{-14}$. Iida, Minakata and Yasuda \cite{IMY} 
discussed the possibility of doing the similar test by the 
long-baseline accelerator neutrino experiments and obtained 
$|\Delta f|\sim10^{-14}$ as expected sensitivity. It is accidental 
that two methods yield the same order of magnitude for the 
sensitivity since the ingredients involved in these estimations
are entirely different with each other. 
Some related considerations were given in Ref. \cite{PHL}.

If one want to use solar neutrinos as an experimental mean to test 
the equivalence principle one has to decide his attitude to  the 
solar neutrino problem \cite{BAH}. One might say that it is due to 
the fact that our ability of modeling the sun is so poor and the 
problem may be solved by inventing an appropriate non-standard 
solar model. Others may argue that it implies a new physics of 
neutrinos beyond the standard electroweak gauge theory. 
Someone would be more brave and may propose 
that it could be explained by the effect of violation of the 
equivalence principle itself, as has been done in Ref. \cite{PHL}.

In this paper we make an assumption that the 
Mikheyev-Smirnov-Wolfenstein (MSW) mechanism \cite{MSW} is the 
cause of the solar neutrino deficit. 
It makes our analysis model-dependent but it is inevitable if the 
solution to the solar neutrino problem involves the physics 
beyond the standard model of electroweak interactions. 
In view of the progress in solar neutrino observation done and to 
be done by the present and the future detectors it would not be too 
optimistic to speculate that we would finally find the answer to 
the solar neutrino puzzle. The point we want to stress is that 
once we have {\it the} solution one can perform similar
analyses as ours to test the equivalence principle and the present 
work may serve as a prototype for them.

In this paper we will make a substantial improvement in the 
treatment of possible experimental test of the equivalence 
principle by solar neutrino observation. A part of this work 
has been presented in Ref. \cite{Mina} though in a 
very preliminary stage.

The points we are going to make in this paper are:

\noindent(1) We shall develop a new formalism which describes 
neutrino propagation in an arbitrary spherically symmetric 
metric of a perfect-fluid star. 
It enables us to calculate neutrino flavor conversion 
probabilities 
that are accurate to order $G$, the Newton constant.

\noindent(2) Our analysis of the effects of violation of the 
equivalence principle relies solely on the spectral shape of 
the $^8\mbox{B}$ neutrinos. 
We fully treat the effects of the MSW transformation due to the 
neutrino flavor mixing as well as that of the non-universal 
gravitational couplings. 
We discuss how one can separate the latter effect from the 
former's and we estimate the sensitivity by fully  taking into 
account of this mixture of two different flavor transformations.

First let us recall the definition of the equivalence principle. 
According to the text book by Misner, Thorn and Wheeler \cite{Grav}, 
the equivalence principle is stated as ``In any and every local 
Lorentz frame, anywhere and anytime in the universe, all the 
(non-gravitational) laws of physics must take on their familiar 
special-relativistic forms.'' 
It follows from the principle that if the gravitational couplings
of different flavor neutrinos are different the equivalence 
principle is violated. 
This is because in a local Lorentz frame where electron
neutrinos obey the familiar special-relativistic wave equation 
muon neutrinos do not.

In this paper we work with two neutrino flavors and denote them 
as electron and muon neutrinos. We consider the Lagrangian
\begin{eqnarray}
{\cal L} &=& 
\sum_{i=1,2} e(G_i)\bar{\nu}_{G_i}ie^{a\mu}(G_i)
\gamma_aD_\mu(G_i)\nu_{G_i} \cr
& & - \sum_{i=1,2} e(G_i)m_i\bar{\nu}_{Mi}\nu_{Mi} \cr
& & + (interactions \, with \, electroweak \, gauge \, fields),
\label{eqn:Lagrangian}
\end{eqnarray}
where $e^{a\mu}(G_i) (i=1,2)$ are the vierbein fields of some 
background metric and $e(G_i) \equiv \mbox{det} e^{a\mu}(G_i)$.
The Newton constants $G_i(i=1,2)$ are allowed to be different
for neutrino basis $\nu_{G1}$ and $\nu_{G2}$.
The last term in (\ref{eqn:Lagrangian}) is assumed to be 
written by flavor (or gauge) eigenstate. We take the most general 
ansatz that the flavor, the gravity and the mass eigenstates are 
all different with each other. 
The latter two are related with the flavor eigenstate in the 
following way:

\begin{eqnarray}
\left[\matrix{\nu_e \cr\    \nu_{\mu} \cr}\right] 
&=&
\left[\matrix{
\cos\theta_G & \sin\theta_G \cr
-\sin\theta_G & \cos\theta_G \cr}
\right]
\left[\matrix{\nu_{G1} \cr\    \nu_{G2} \cr}\right]\nonumber,\\
&=&
\left[\matrix{
\cos\theta_M & \sin\theta_M \cr
-\sin\theta_M & \cos\theta_M \cr}
\right]
\left[\matrix{\nu_{M1} \cr\    \nu_{M2} \cr}\right].
\label{eqn:matrix}
\end{eqnarray}
The mixing angle $\theta_M$ in (\ref{eqn:matrix}) is the usual 
flavor mixing angle which plays an important role in the 
MSW mechanism.

As a measure for violation of the equivalence principle we define 
\begin{equation}
\label{eqn:violation}
\Delta f = {G_2 - G_1   \over {1  \over 2}(G_2 + G_1)}.
\end{equation}
We believe that $\Delta f$ defined in (\ref{eqn:violation}) does 
have a right correspondence with $\eta$ which is defined in a 
similar fashion as
(\ref{eqn:violation}) by replacing $G_i$ by $(M/m)i$ 
as a measure for the in-equivalence of inertial $(m)$ to 
gravitational $(M)$ masses. 
The equivalence of $\eta$ and $\Delta f$ comes closer to be true 
if we are allowed to boldly interpret the E\"otv\"os-Dicke-type 
experiments as measuring the non-universality of gravitational 
couplings of u- and d-quarks.

We shall derive the neutrino evolution equation under the 
influence of a weak gravitational field of the sun. For generality 
we work with the following most general static and spherically 
symmetric metric, 
\begin{eqnarray}
ds^2 = g_{tt}(r) dt^2 + g_{rr}(r) dr^2 - 
r^2( d\theta^2 + \sin^2\theta d \phi^2),
\label{eqn:metric}
\end{eqnarray}
taking the center of the sun as the origin of the coordinate. 
We make the ansatz that the energy-momentum tensor of 
the solar matter is given by the form of perfect fluid, 
\begin{equation}
\label{eqn:tensor}
T_{\mu\nu} = P g_{\mu\nu} + (P + \rho) U_\mu U_\nu,
\end{equation}
which should give 
an excellent approximation to the sun.
In (\ref{eqn:tensor}) $\rho$ and $P$ denotes the energy density 
and the pressure, respectively, and $U^{\mu}$ is the usual 
velocity four vector.
The metric  $g_{rr}(r)$ and $g_{tt}(r)$ 
can be obtained by solving the Einstein equation and are given 
up to the order of $G$ by \cite{Grav}
\begin{eqnarray}
\label{eqn:vierbein}
g_{tt}(r) &=& 
1-2 G\  \int_{ r}^{ \infty} \left[ {{\cal M}(r') \over r'^2}
+ 4 \pi r' P(r')\right] dr',  \\
-g_{rr}(r) &=&1+2 G\ {{\cal M}(r) \over r},
\end{eqnarray}
where

\begin{equation}
{\cal M}(r) = \int_0^r dr' 4\pi r'^2 \rho(r')
\label{eqn:calM}
\end{equation}
and it can be interpreted as the total mass contained in the 
volume with radius $r$ for Newtonian stars.
In case of the sun, we can neglect the pressure term in equation 
(\ref{eqn:vierbein}) which is very small compared to the 
term ${\cal M}(r)/r^2$.
In fact one can show that the pressure term is of order $G^2$ 
by noticing the equation of hydrostatic equilibrium, 
$r^2[ dP(r)/dr] = -G {\cal M}(r) \rho$.
This equation holds in most of the stars for which 
the Newtonian approximation is valid.
We cannot neglect the pressure term in such stars as 
white dwarfs or neutron stars. We will 
neglect the pressure term in the following discussions. 

Here we give some clarifying comments on the theoretical basis and 
the assumptions we make in 
doing pilot computations for testing the equivalence 
principle by using solar neutrinos. 
First of all we ignore the effects of
gravitational field of the earth, where the neutrino detector 
is located, by assuming that the sun's gravitational field gives 
a dominant effect. 
Strictly speaking we assume in our treatment that the 
detector is located at far apart from the 
sun where the metric is approximately Minkowskian. 
We also ignore the possible MSW transformation in the earth which 
may not give a good approximation with the large-mixing-angle 
solution of the solar 
neutrino problem. 

One of the most important issue which remains to be explored is the 
problem of possible frame-dependence in the result. 
We shall make some comments on it while it will be addressed in more
detail in Ref.  \cite{IMY2}.
Since the general coordinate invariance is broken in the 
theory defined by (1) there is no guarantee that 
the bound we obtain is independent of the frame we choose. 
Namely, we will do our computation by taking the inertial frame 
of the sun, but the result of neutrino flavor transformation 
might be different if we do the computation 
in the inertial frame of the earth even if we keep ignoring 
the effect of earth's gravitational field.

We argue, however, that the corrections due to 
the frame-dependence is of the order of $\Delta f$, and 
is negligibly small if we restrict ourselves into, 
for example, $| \Delta f | \le 10^{-10}$. 
To this goal let us assume that by taking 
limit of $\Delta f \rightarrow 0$ the theory defined by (1) 
approaches to general relativity coupled with matter-gauge fields.
We are to deal with weak gravitational fields of the sun
and confine ourselves into the perturbation expansion 
up to first order in $G$. In this 
weak-field approximation the above assumption should be true. 
Then in the limit of  $\Delta f \rightarrow 0$ there
should be no frame dependence as it is the general 
relativity. 
In fact, it is a Newtonian gravity with special relativity.  
We conclude that within the weak-field approximation the frame 
dependence comes in as the first-order in $\Delta f$
and would leave only negligible effects in observables. 
In particular the coefficient of $\Delta f$ is frame 
independent.

Let us consider the case that the electron 
neutrinos are produced at the origin 
and propagate toward the  solar surface.  In this case 
we can choose the coordinate with $\theta = \phi = 0$ 
to describe the trajectory of neutrino propagation. 
Under this choice of the coordinate, the Dirac equation is given by
\begin{equation}
[i(e^0_t)^{-1}\gamma_0 \partial_t+i(e^1_\gamma)^{-1}
\gamma^1\partial_r-m]\psi = 0,
\label{eqn:Dirac}
\end{equation}
where we note that the vierbeins  
$e^0_t$ and $e^1_r$ are  
given by the metric as 
$(e^0_t)^2 = g_{tt}$ and $(e^1_r)^2 = g_{rr}$.
Here we have ignored the derivative terms in spin connections 
because it is of the order of 
$\sim 1/ER_\odot \sim 10^{-22}$ for 
E=10 MeV and is negligibly small.
 
By taking into account only the positive energy component of 
the Dirac equation and choosing outgoing wave, we 
get the evolution equation in the weak gravitational field  of 
the sun for one generation of neutrino,
\begin{equation}
i{d \over dr} \nu 
= E \left[1- {m^2 \over 2E^2} 
-\phi(r)
\right]\nu,
\end{equation}
where
\begin{equation}
\phi(r) \equiv
 - G 
\left[ {{\cal M}(r) \over r}+ \int_{ r}^{ \infty}
 {{\cal M}(r) \over r^2} dr \right].
\end{equation}

It is easy to generalize this equation 
to the two flavor case described by the Lagrangian 
(\ref{eqn:Lagrangian}). The evolution equation in the sun for 
two neutrino flavors is given by 

\begin{equation}
\kern-1.0in
i{d \over dr}\left[\matrix{\nu_e \cr\    \nu_{\mu} \cr}\right]
=H\left[\matrix{\nu_e \cr\    \nu_{\mu} \cr}\right],
\label{eqn:evolution}
\end{equation}
where 

\begin{equation}
H
=\left[\matrix{ &\kern-0.15in
 \sqrt{2} G_F N_e +
{\Delta_M} \sin^2\theta_M +
{\Delta_G} \sin^2\theta_G 
  & {1 \over 2} ({\Delta_M } \sin2\theta_M+
  {\Delta_G} \sin2\theta_G )\cr
  &{1 \over 2}({\Delta_M} \sin2\theta_M +\Delta_G \sin2\theta_G )
  &{\Delta_M} \cos^2\theta_M  + {\Delta_G} \cos^2\theta_G \cr } 
\right] 
\label{eqn:hamiltonian}
\end{equation}
with 
\begin{eqnarray}
\Delta_M &\equiv \Delta m^2/2E,\\
\Delta_G &\equiv \Delta f \phi(r) E.
\end{eqnarray}
where $\Delta m^2 = m_2^2 - m_1^2.$
Note the difference in energy dependences of $\Delta_M$ and 
$\Delta_G$. 
This makes it possible to distinguish the effect of 
the violation of equivalence principle from that of the MSW effect 
by examining modulation of the neutrino energy spectrum.

With one more ingredient having added to the MSW mechanism 
the resultant equation of neutrino transformation 
(\ref{eqn:evolution}) is far richer than the usual MSW's. 
We only make a few remarks on some distinctive features of it. 
The resonance condition is obtained by equating the two diagonal 
terms in the Hamiltonian matrix (\ref{eqn:hamiltonian}), 
\begin{equation}
\sqrt{2} G_F N_e -{\Delta_M} \cos2\theta_M -{\Delta_G} 
\cos2\theta_G =0
\label{eqn:condition}
\end{equation}
{}From (\ref{eqn:condition}) one observes several notable features:
\noindent(1) The resonance condition can be satisfied even for 
$\Delta m^2 = 0$ if $\Delta f < 0$, the gravitational MSW effect.

\noindent(2) Resonance can occur at outside the sun where 
$N_e = 0$ if $\Delta m^2\cdot \Delta f > 0$.
This should not come as a surprise because $\phi(r)$ plays 
the role analogous to the position-dependent electron density 
in the MSW effect.

\noindent(3) The mathematical structure of the equation 
(\ref{eqn:evolution}) is quite different from that of the flavor
MSW mechanism by having $r$-dependent term in the off-diagonal 
element in the Hamiltonian (\ref{eqn:hamiltonian}). 
This makes the usual Landau-Zener analysis difficult, or at least 
highly nontrivial.

Notice that not only the condition 
(\ref{eqn:condition}) but also 
the adiabaticity 
condition is needed for sufficient resonant conversion of 
neutrinos. 
We can obtain the adiabaticity condition by demanding that the 
oscillation length at a resonance point is much shorter 
than the width of the resonance region. 
For resonance in the interior of the sun the adiabaticity 
condition reads, 
\begin{equation}\left.
{h \over 2\pi} {(\Delta_M \sin2\theta_M +  
\Delta_G \sin2\theta_G)^2 \over
(\Delta_M\cos2\theta_M + {\widetilde \Delta_G} \cos2\theta_G)}
\right|_{r=r_{res}} \gg 1
\label{eqn:adiabatic}
,\end{equation}
where 
\begin{equation}
{\widetilde \Delta_G} \equiv \Delta_G + h (d\Delta_G /dr).
\end{equation}
To express the adiabaticity condition in the form of 
(\ref{eqn:adiabatic}) 
we have approximated the electron number density $N_e$ 
in the sun as
\begin{equation}
N_e(r) = N_0 \exp(-r/h)\hskip 1cm  (r>0.1 r_\odot) .
\end{equation}
with $h\sim r_\odot/10$.

Now we will discuss how sensitively violation of the equivalence 
principle affect the solar neutrino observation. 
To compute ${P(\nu_e\rightarrow \nu_x; E_\nu)}$, 
the probability of observing $\nu_x$ ($x=e$ and $\tau$) at 
the earth, we employ the following procedure. 
We numerically integrate the evolution equation 
(\ref{eqn:evolution}) from $r=0$ to $r = 10r_\odot$ where 
the gravity effect is 
negligible, $|\Delta_G |<< \Delta_M$, for 
regions of parameters of interest to be specified below. 
We ignore the term $\Delta_G$ at $r > 10r_\odot$ and 
evaluate   ${P(\nu_e\rightarrow \nu_x; E_\nu)}$ by 
\begin{eqnarray}
\label{eqn:average}
{P(\nu_e\rightarrow \nu_e; E_\nu)} &=& 
\left(1-{1\over2}\sin^22\theta_M\right) |A_{\nu_e}|^2
+{1\over2}\sin^22\theta_M \  |A_{\nu_\mu}|^2
\nonumber\\
&&
 -\sin 2\theta_M \cos 2\theta_M\ {\rm Re}\left[
A_{\nu_e}A_{\nu_\mu}^*
\right]\\
{P(\nu_e\rightarrow \nu_\mu; E_\nu)}&=& 
1-{P(\nu_e\rightarrow \nu_e; E_\nu)} 
\end{eqnarray}
where $A_{\nu_e}$ and  $A_{\nu_\mu}$ denote the 
probability amplitudes of neutrinos being $\nu_e$ 
and $\nu_\mu$ at $r=10r_\odot$, respectively. 

We calculate the electron energy spectrum $f(E_e)$ 
to be measured at the Super-Kamiokande by the following equations
\begin{eqnarray}
\label{eqn:Kamiokande}
f(E_e) &=& 
 \sum_{x=e,\mu} 
\int\limits_0^{\infty}dE_\nu
\ \varepsilon(E_e) {[d \sigma_{\nu_xe}(E_\nu,E_e)/ dE_e]}
\makebox[2cm]{}
\nonumber\\
&& \makebox[2cm]{}\times{P(\nu_e\rightarrow \nu_x; E_\nu)}
{[d \phi_{^8{\rm B}}(E_\nu)/ dE_\nu]},
\end{eqnarray}
where 
${d \sigma_{\nu_xe}(E_\nu,E_e)/ dE_e}(x=e,\mu)$  are the 
neutrino-electron elastic scattering cross section 
and ${d \phi_{^8{\rm B}}(E_\nu)/ dE_\nu} $
is the differential flux of $^8{\rm B}$ neutrino 
calculated by the standard solar model (SSM)  \cite{BU}. 

For the Sudbury Neutrino Observatory (SNO) detector we use 
a similar expression as (\ref{eqn:Kamiokande}), the one in 
which one restricts 
the summation over flavors into electron and replaces 
the neutrino-electron scattering cross section by the electron 
energy distribution for a given neutrino energy, 
$dD(E_\nu, E_e)/dE_e$, for the reaction 
$\nu_e + d \rightarrow p + p + e^-$ 
calculated by Nozawa  \cite{Nozawa}.
It includes the effects of (a) the Fermi motions of nucleons in the 
target deuterons, (b) the final state three-body kinematics, 
and (c) the final state Coulomb interactions.
We have used very simplified form of 
the detection efficiency $\varepsilon(E_e)$ as 
\begin{equation}
\varepsilon(E_e) = \theta(E_e-E_{th}),
\label{eqn:Super}
\end{equation}
with $E_{th} = 5 $ MeV for both of the detectors.
Alternative forms of (\ref{eqn:Super})  may not alter our conclusion 
because the effect of gravity exists in high energy part of 
the neutrino spectrum. 
We have taken into account the expected energy resolution 
$14\%/ \sqrt{E/10\mbox{MeV}}$ for Super-Kamiokande  \cite{Tot} and 
$10\%/ \sqrt{E/10\mbox{MeV}}$ for SNO  \cite{Hata}.

As a prototype analysis in this paper we pick up the two sets of 
parameters of flavor mixing as representatives of (i) the 
nonadiabatic and 
(ii) the large-mixing-angle MSW solutions: 
(i) $\Delta m^2 = 6\times 10^{-6}{\rm eV}^2$, 
$\sin^2 2\theta_M = 0.01$ and 
(ii) $\Delta m^2 = 2\times 10^{-5}{\rm eV}^2$, 
$\sin^2 2\theta_M = 0.7$.

We do not make any attempt to average over the production points of 
$^8$B neutrinos but assume that they are produced at the center 
of the sun. Given the parameter of interest the resonance 
points are well outside the production region of $^8$B neutrino
and therefore our treatment should provide a reasonably good 
approximation.

To have a feeling on how sensitively violation of the 
equivalence principle affect the MSW-modulated neutrino 
spectrum we perform a pilot computation by taking $\theta_G$ 
and $\theta_M$ equal for the case of nonadiabatic 
MSW parameters (i). In Figs. 1 and 2 we show 
(a) recoil electron energy spectra, and (b) electron 
energy spectrum divided by the one predicted by SSM. 
Fig. 1 is for Super-Kamiokande and Fig. 2 is for SNO.
We have done the computation 
with five parameters, $\Delta f=0, \pm5.0\times 10^{-16},$ and 
$\pm1.0 \times 10^{-15}$. In Figs. 1(b) and 2(b) the ratios are 
normalized at 9 MeV of electron energy.
Super-Kamiokande is expected  \cite{Tot} to have accuracy of 2-3\% 
for the ratio of modulated spectrum to that of SSM 
after its 3 years operation depending 
on the energy region and the MSW parameters . 
Therefore, from Fig. 1 we expect that the effect of 
$|\Delta f|$ of the order of $10^{-15}$ can be 
distinguished from the pure 
nonadiabatic MSW solution.  

The similar computation can be done with the parameters 
(ii) of the large-mixing-angle MSW solution. But we do not 
show the result here for reasons which will be explained 
after the full analysis with the 
large-mixing angle MSW solution.

A technical comment is in order.
Because of the limitation of the computer time we cut 
off the lower end of integration over the neutrino energy in 
(\ref{eqn:Kamiokande}) at $E_\nu$ = 5 MeV.
Since we take into account the energy resolution of about 1 MeV  
(Super-Kamiokande) and 0.7 MeV (SNO) at electron energy 
of 5 MeV we would have to include the contribution from neutrino 
energy $E_\nu \gsim 3$ MeV.  
It thus provides
an unconventional way of smoothing out the effect of 
the sharp (step function) cut-off (\ref{eqn:Super}) 
of the detection efficiency.  
We believe that it does not gives rise to any serious effects to
our analysis because we use the ratio of the energy spectrum 
to that of SSM. 

Now we enter into a full analysis with the nonadiabatic MSW 
solution of the solar neutrino problem. 
We aim at identifying the region of 
parameters $\Delta f$ and $\sin 2\theta_G$ in which the spectral 
shape of $^8 B$ neutrinos can be distinguished 
with 90\% confidence 
level from that predicted by the pure MSW solution with the 
particular set of parameters (i) and (ii). 
At the present stage it is of course 
impossible to pinpoint the MSW parameters even if we assume 
that it is {\it the} solution to the solar neutrino problem. 
The present work with the  particular set of parameters, 
therefore, is only meant to present a proto-typical analysis 
toward the more complete ones. 

We define, as a quantitative measure for the deviation 
of the electron energy spectrum,
\begin{equation}
\chi^2 = {\rm min} \left[
\sum_{i=1}^N 
\left[\frac{x_i(\Delta f=0)-\alpha x_i(\Delta f,\sin2\theta_G)}
{\sigma_i({\rm st})}\right]^2 
+ \left({1-\alpha \over \sigma_i({\rm sys})}\right)^2
\right],
\label{eqn:spectra}
\end{equation}
following the procedure of Ref.  \cite{Kamii}.
We divide the energy range into $N$ = 19 bins and 
calculate the count rate by 
integrating the electron energy spectrum in each bin. 
$x_i$ in (\ref{eqn:spectra}) implies the ratio of 
the count rate calculated in this way to 
the one calculated in the same way but using 
SSM without any effects of flavor-gravitational transformation. 
$\sigma_i({\rm st})$ and $\sigma_i({\rm sys})$ denote
the statistical and the systematic errors, 
respectively. 
The symbol $min$ in front of the right-hand-side of 
(\ref{eqn:spectra}) is 
to take minimization by varying the parameter $\alpha$. 
The measure $\chi^2$ 
is thereby sensitive only to the shape of the energy spectra
within the uncertainty of absolute normalization 
represented by the systematic error. 

In our computation we treat the statistical 
error in the following way. We suppose 
that the Super-Kamiokande (SNO) detector 
observed 20000 (6000) and 40000  (12000) solar 
neutrino events, which roughly 
corresponds to its operation over 2 and 4 years, respectively.
Then, it is rather straightforward to evaluate the statistical 
error by computing number of events in each bin. 
We combine the errors of the numerator and 
of the denominator by assuming that 
they are independent Gaussian errors. 
Lacking detailed knowledge of the detector performance 
we simply assume, as systematic errors of both the detectors, 
$\sigma_i({\rm sys}) =\sigma({\rm sys})$ = 6\%, 
the systematic error of the Kamiokande-II experiment  \cite{Kamii}.

We emphasize that our treatment of errors does not have any 
particular significance and should be taken only as tentative. 
In particular we did not attempt to make a detailed comparison 
between the expected sensitivities between the two detectors. 

In Fig. 3 we present the results of the analysis for 
Super-Kamiokande in cases of (a) $\Delta f>0$ and 
(b) $\Delta f<0$. Plotted in Fig. 3 are the regions of 
parameters on $\Delta f-\sin 2\theta_G$ plane which are 
sensitive at 90\% confidence level to the effect of violation 
of equivalence principle. Notice that the relative sign 
of $\theta_G$ with respect 
to $\theta_M$ is physically meaningful. 
The open circles and the asterisks are for cases with 40000 
and 20000 events, respectively. 
We observe that the sensitive region goes down to 
$|\Delta f|=10^{-15}-10^{-16}$ depending upon $\sin2\theta_G$.
It is notable that a factor of 2 better statistics achieved 
by taking the data for 2 more years does not significantly 
improve the sensitivity.  It is largely due to a sharp growth 
of chi-square near the edge of the sensitivity region. 

The same analysis is repeated for the SNO detection and 
the results are  presented in Fig. 4. The notations are exactly 
the same as in Fig. 3. In spite of the assumed factor of 
$\sim \sqrt{3}$ larger statistical errors the sensitivity 
of SNO to the violation of the equivalence principle is 
essentially equal to that of Super-Kamiokande.
It is due to an advantage of the deuterium detector 
in which one can probe neutrino spectrum more accurately 
than the light-water detector by use of the absorption 
reaction $\nu_e+d \rightarrow p+p+e^-$. 
If we ignore the three effects mentioned before the electron 
energy is equal to the neutrino energy minus 1.44MeV. 
On the other hand the measurement of the recoil electron 
energy implies taking convolution (smearing) of the original 
neutrino spectrum. 

In Fig. 5 and 6 we present the results with the parameter 
(ii) which corresponds to the large-mixing-angle solution. 
Fig. 5 is for Super-Kamiokande and Fig. 6 is for SNO. 
Again the results are almost the same.
It is evident that the sensitivity is at most 
$\Delta f \sim 10^{-14}$ at some particular values of 
$\sin 2\theta_G$ and in general $|\Delta f| \gsim 10^{-13}$.

What is the reason for such (relative) great insensitivity in 
this case? 
In what follows we give a very simple-minded argument which 
apparently explains this behavior.
For the parameter set (ii) with $\Delta f \sim  10^{-14}$, 
the adiabatic condition is almost completely satisfied except for 
the case  
$\theta_M = \theta_G$. 
An angle which diagonalizes the 
Hamiltonian matrix (\ref{eqn:hamiltonian}) is given by
\begin{equation}
\sin2 \theta = \frac{\Delta_M S_M+\Delta_G S_G}
{\sqrt{(\Delta_M S_M+\Delta_G S_G)^2+
(\Delta_M C_M+\Delta_G C_G-\sqrt{2} G_F N_e )^2}},
\label{eqn:angle}
\end{equation}
where $S_{M,G} \equiv \sin2\theta_{M,G}$ and 
$C_{M,G} \equiv \cos2\theta_{M,G}$.
By using this $\theta$, the two energy eigenstate $\nu_1$ and 
$\nu_2$ of 
$H$ is given by 
\begin{equation}
\left[\matrix{\nu_1 \cr \nu_2}\right] =
\left[\matrix{\cos\theta & -\sin\theta\cr 
\sin\theta&\cos\theta}\right]
\left[\matrix{\nu_{e}\cr \nu_{\mu}}\right], 
\label{eqn:mixing}
\end{equation} 
At the center of the sun,  $\theta \sim \pi/2$ because the 
matter effect $N_e$ dominates over the mass term and the 
gravitational 
term (see eq. (\ref{eqn:angle})). This means that $\nu_e$ 
produced in the center of the sun is essentially equal to the 
state $\nu_2$ (see eq. ({\ref{eqn:mixing})).
On the other hand, at the earth $\theta \sim \theta_M$ because 
there is no matter effect ($N_e = 0$) and $\Delta_G \ll \Delta_M$ 
for the parameter set (ii) with $\Delta f \sim 10^{-14}$.
Hence  if the resonance occurs adiabatically the probability of 
observing $\nu_e$ at the earth is given by 
\begin{equation}
P(\nu_e) \sim | \langle{\nu_e}| \nu_2 \rangle|_{earth}^2 = 
\sin^2\theta_{earth}\sim\sin^2\theta_M. 
\label{eqn:earth}
\end{equation}
Of course this is the same expression as we obtain in 
the large mixing angle MSW mechanism by 
which the flux of 
solar neutrinos is uniformly reduced independent of energy. 
Thus we expect 
the non-universality of the neutrino 
gravitational 
couplings is unlikely to produce detectable effect in modulation 
of the neutrino energy spectrum with $|\Delta f|$ smaller than 
$10^{-14}$, if adiabaticity condition holds. 
Far better sensitivity with the parameter (i) is then understood 
to be due to the non-adiabaticity of the neutrino 
flavor-gravitational transformation. 

If we increase the value of $|\Delta f|$ beyond $\sim 10^{-13}$ 
the sensitivity to the non-universal gravitational coupling of 
neutrinos depends on the sign of $\Delta f$. 
As we see in Fig. 5 and 6 it produces an appreciable 
difference with the large-mixing-angle parameter (ii).
Let us examine negative $\Delta f$ case first 
to understand the reasons for the difference. 
In view of the resonance condition (\ref{eqn:condition})
the resonance ceases to occur if $\Delta f <0$ and 
$|\Delta f| \gsim 10^{-13}$ because the gravity effect 
(third term of  (\ref{eqn:condition})) tends to dominate over the 
matter effect (first term). If the resonance does not occur 
the spectrum of course deviates from the one 
expected by the MSW mechanism. It explains the region 
of sensitivity which starts to develop at 
$-\Delta f \simeq 5\times 10^{-14}$ in Figs. 5b and 6b.

In case of positive $\Delta f$ the situation is quite different. 
If $\Delta f \gsim 10^{-14}$ the resonance can not occur 
inside the sun because both the first 
and the third terms of  (\ref{eqn:condition}) give 
positive contributions. So 
the resonance occurs between the sun and the earth. For this 
resonance the adiabaticity condition is fulfilled and 
the argument leading to   (\ref{eqn:earth}) holds.
The insensitivity exhibited in Figs. 5a and 6a follows
because of the difficulty in perturbing the adiabatic 
large-mixing-angle mechanism by the effect of $\Delta f$. 
We should note that our computation is not very accurate in 
regions $|\Delta f| \gsim 10^{-14}$ for these particular 
figures because we only integrate the equation 
(\ref{eqn:evolution}) up to $r = 10r_\odot$.

In the analysis in this paper we have assumed that the MSW 
mechanism is the cause of the solar neutrino deficit. 
How reliable is this assumption in the light of 
existing data of ongoing four experiments? An extensive 
analysis by Hata and Langacker  \cite{Hata2} seems to 
indicate that the MSW mechanism is the most plausible 
solution to the solar neutrino problem. 
Of course, we have to wait for the  next generation 
experiments, Super-Kamiokande, SNO, and  BOREXINO, to see
if this really is the case. 
At the present stage of solar neutrino experiments 
it is worthwhile to examine the expected sensitivity with 
other solutions to the solar neutrino problem. We hope 
that we will be able to return to this problem in 
the near future. 

In summary we have discussed the possibility of testing the 
equivalence principle by using solar neutrinos. 
We have presented an appropriate formalism for treating neutrino 
propagation under the influence of the weak gravitational 
fields of the sun. 
Assuming the MSW mechanism as the origin of the observed solar 
neutrino deficit we analyzed a complex process of neutrino flavor 
transformation with coexisting effects of the flavor mixing and 
the assumed non-universal gravitational couplings of neutrinos. 
We have obtained, as expected sensitivities, 
$|\Delta f| \sim 10^{-15}-10^{-16}$ 
for a parameter corresponding to the nonadiabatic solution and 
$|\Delta f| \gsim 10^{-13}$ for the large-mixing angle solution. 
We emphasize that it opens a remarkable possibility that solar 
neutrino observation by next generation (light- or heavy-) water 
Cherenkov detectors can improve the experimental bound on 
violation of the principle of equivalence by 3-4 orders of 
magnitude. 

\vskip 2em
\noindent
{\Large{\bf  Acknowledgements}}\\

We are grateful to Satoshi Nozawa for kindly providing us 
with his fortran subroutine for the electron energy 
distribution in SNO, and to Gene Beier and Naoya Hata for their 
informative correspondences. 
We thank Arthur Halprin for informative and critical 
discussions. The works of H. M. and H. N. are partially 
supported by Grant-in-Aid for Scientific Research of 
the Ministry of Education, 
Science and Culture \#05640355, \#05-2115, 
respectively.

We thank Wick Haxton and Institute for Nuclear Theory at 
the University of Washington for their hospitality and the 
Department of Energy for partial support during the completion 
of this work.

\vskip 0.2cm
\newpage
\noindent
{\Large{\bf Figures Captions}}\\
\vskip 0.2cm
\noindent
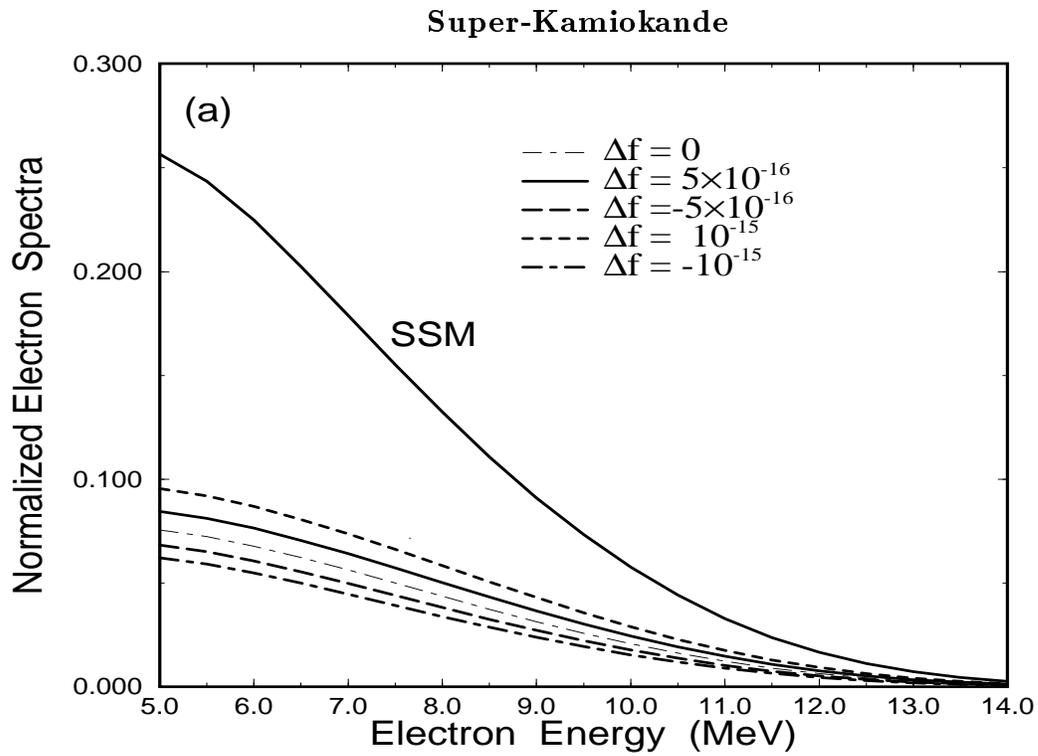
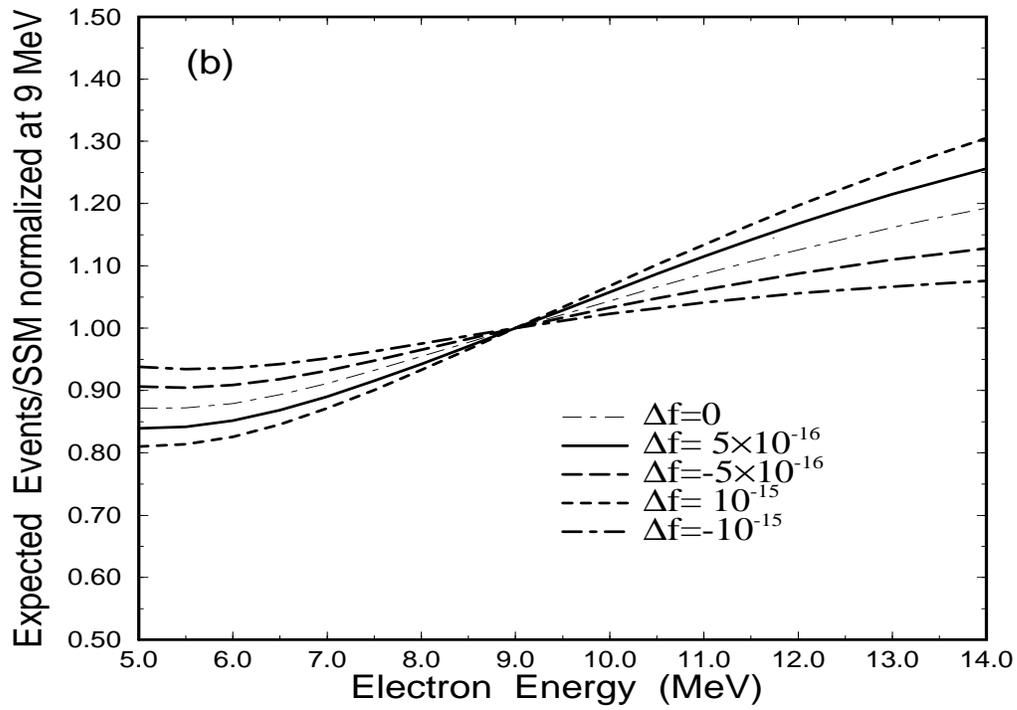
Fig. 1.  Modulation of (a) recoil electron energy spectrum
and (b) the ratio of electron energy spectrum to that of 
SSM to be seen at the Super-Kamiokande detector are indicated. 
The ratios in (b) are normalized 
at $E_e = 9$ MeV. 
The flavor mixing parameters are taken as 
(i) $\Delta m^2 = 6\times 10^{-6}$ eV$^2$ and 
$\sin^2 2\theta_M = 0.01$ 
as a representative of the nonadiabatic MSW solution 
to the solar neutrino problem. 
The presented five lines are for 
$\Delta f = 0, \pm 5.0 \times 10^{-16}$, 
and $\pm 10^{-15}$ with corresponding line symbols shown in the 
figure and $\theta_G$ is taken to be equal to $\theta_M$.
The upper solid line in Fig. 1 (a) which is denoted as 
SSM represents the energy spectrum expected by SSM. 
\vskip 0.3cm 
\noindent
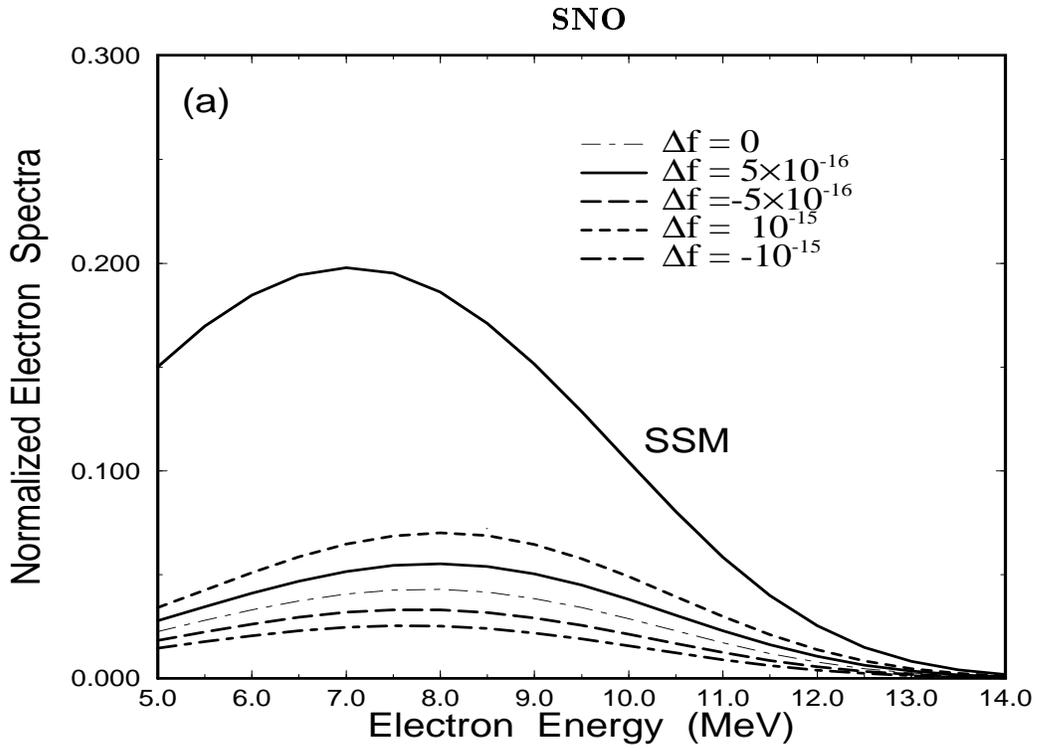
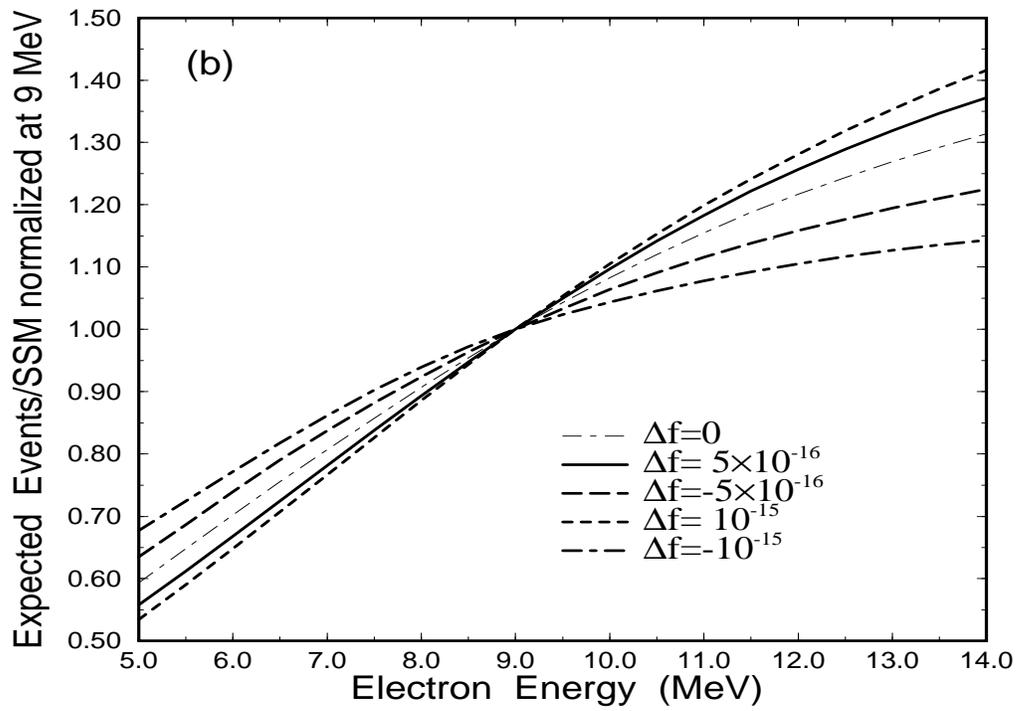
Fig. 2.  The same as in Fig. 1 but for SNO. 
\vskip 0.3cm 
\noindent
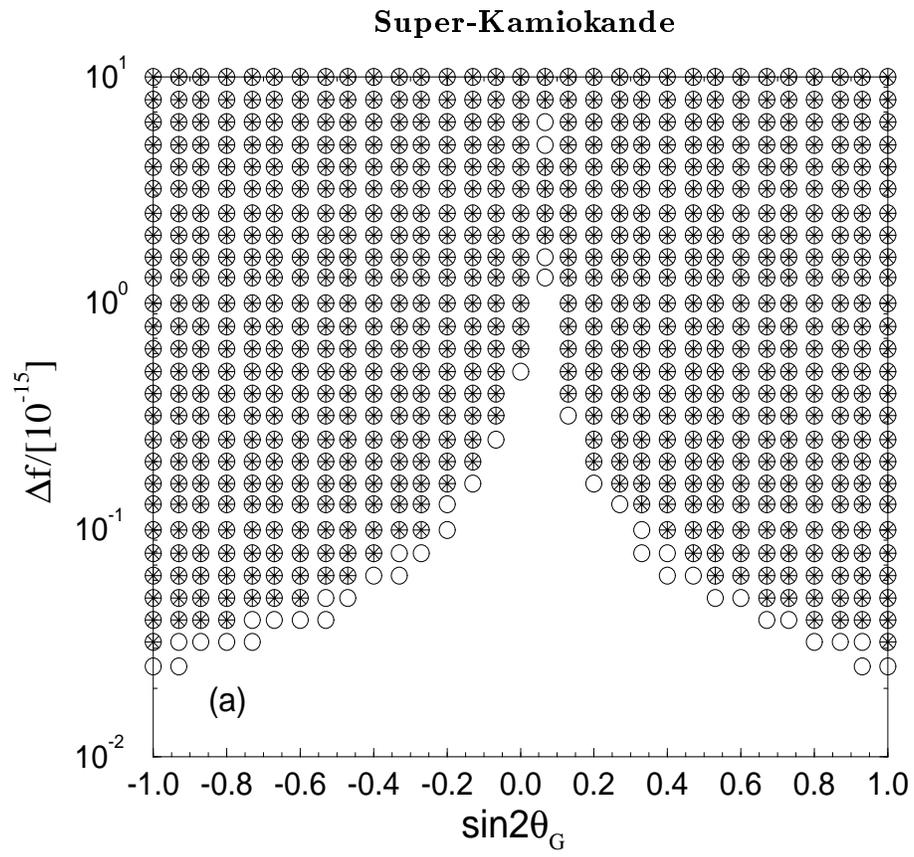
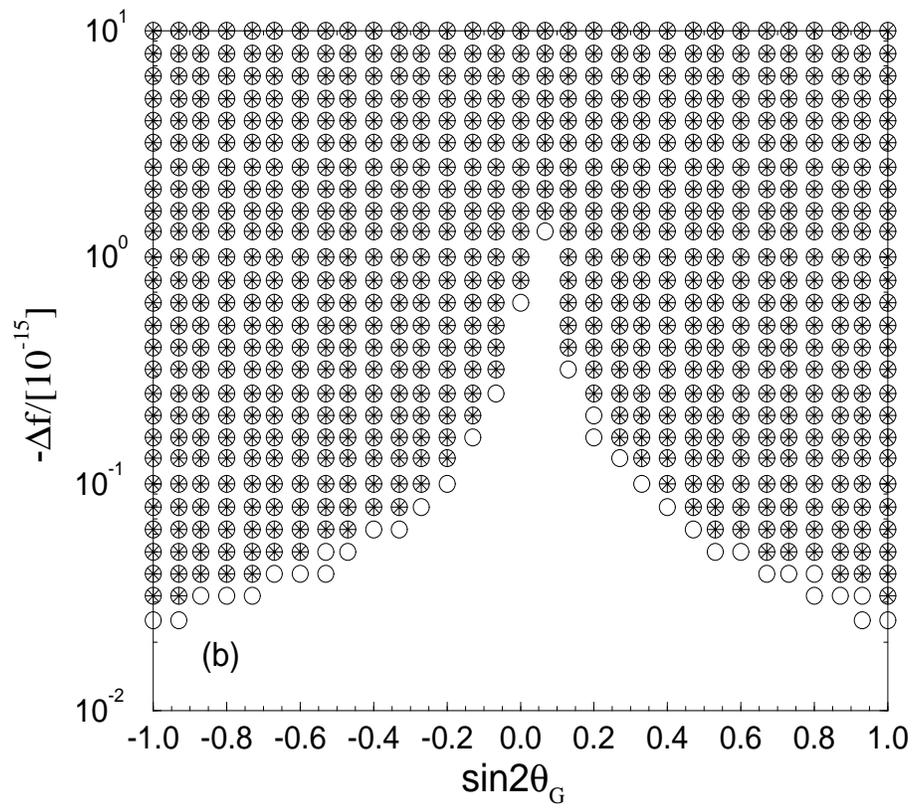
Fig. 3.  The regions  of sensitivity at 90 \% 
confidence level  
to the violation of the equivalence principle 
are drawn on the parameter plane of $\Delta f-\sin 2\theta_G$. 
The result is for Super-Kamiokande and, 40000 
(open circles) and 20000 (asterisks) solar neutrino 
events are assumed, 
roughly corresponding to its operation over 4 and 2 years, 
respectively. 
The flavor mixing parameter for nonadiabatic MSW 
solution are taken as 
(i) $\Delta m^2 = 6\times 10^{-6}$ eV$^2$ and 
$\sin^2 2\theta_M = 0.01$. 
\vskip 0.3cm 
\noindent
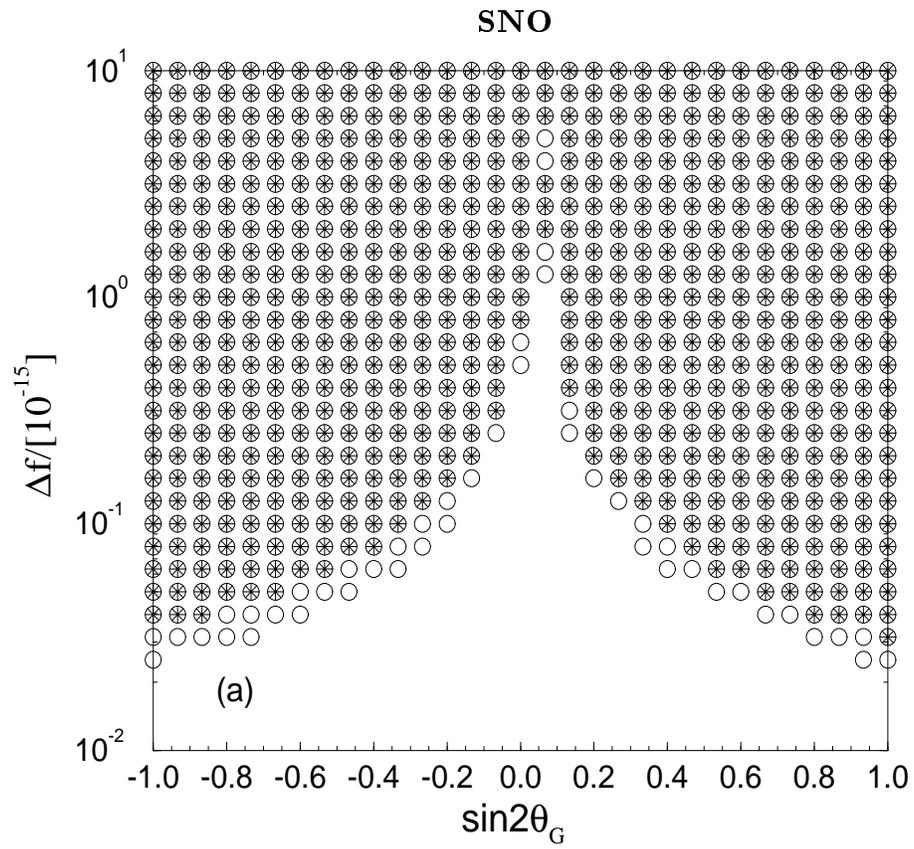
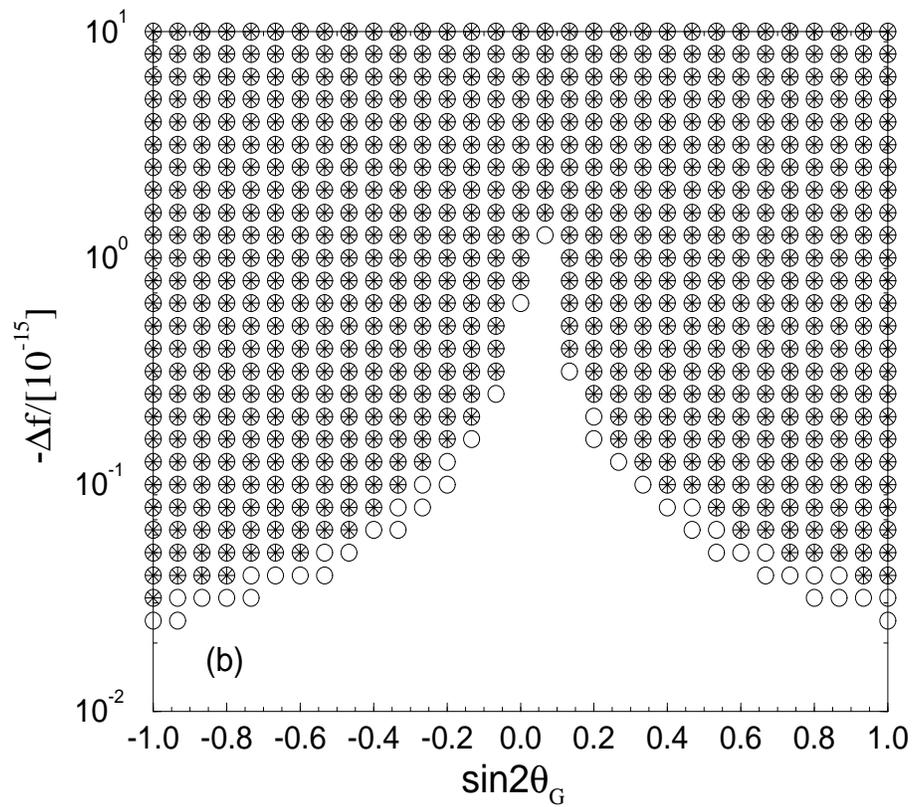
Fig. 4.  The same as in Fig. 3 but for SNO. 
The numbers of events are taken as 
12000  (open circles) and  6000  (asterisks),
roughly corresponding to its operation over 4 and 2 years, 
respectively. 
\vskip 0.3cm 
\noindent
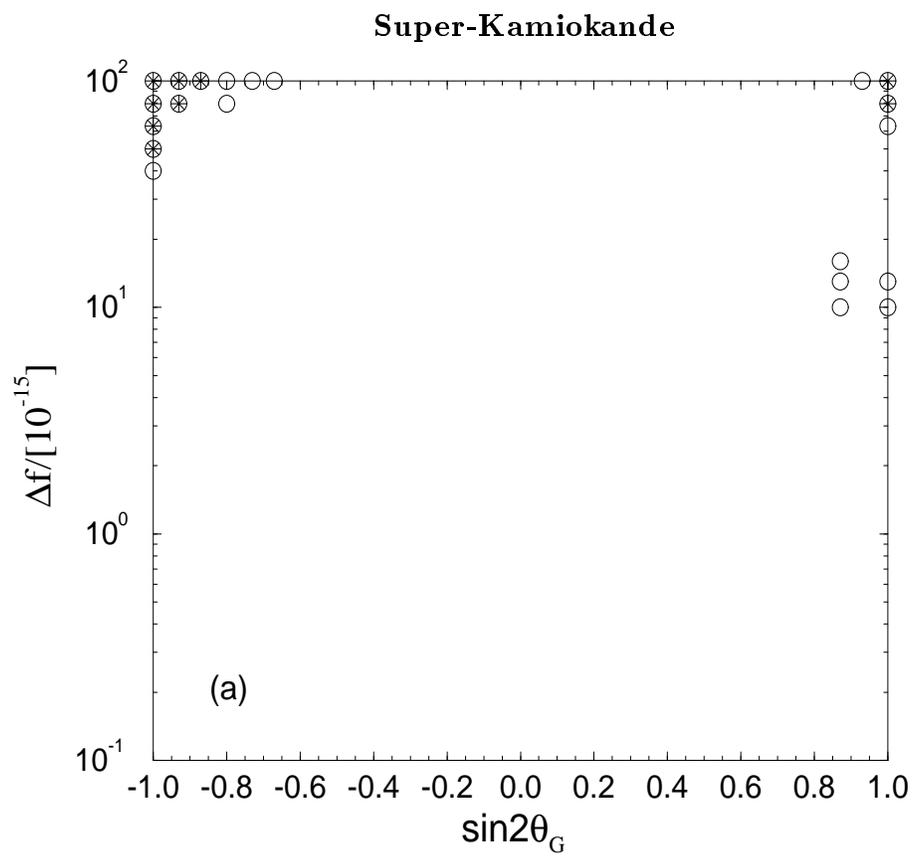
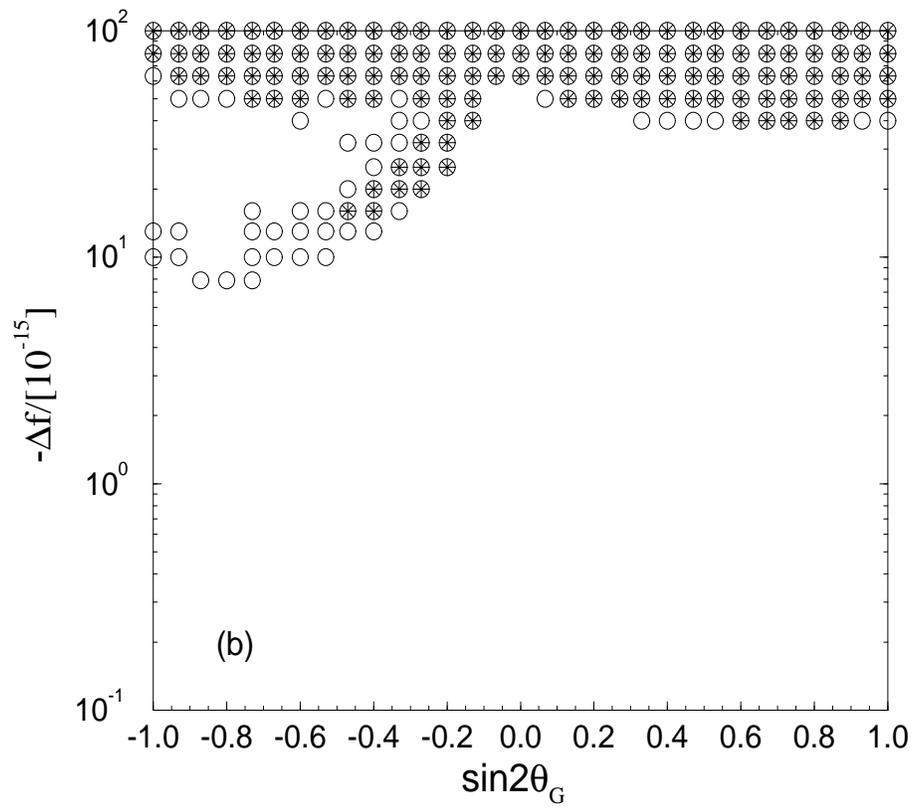
Fig. 5.  The 90 \% confidence level regions of 
sensitivity are drawn for flavor mixing parameter (ii) 
$\Delta m^2 = 2\times 10^{-5}$ eV$^2$ and $\sin^2 2\theta_M = 0.7$ 
of large-mixing angle MSW solution. 
The result is  for Super-Kamiokande and assumes 
40000 (open circles) and 20000 (asterisks) solar neutrino events.
\vskip 0.3cm 
\noindent
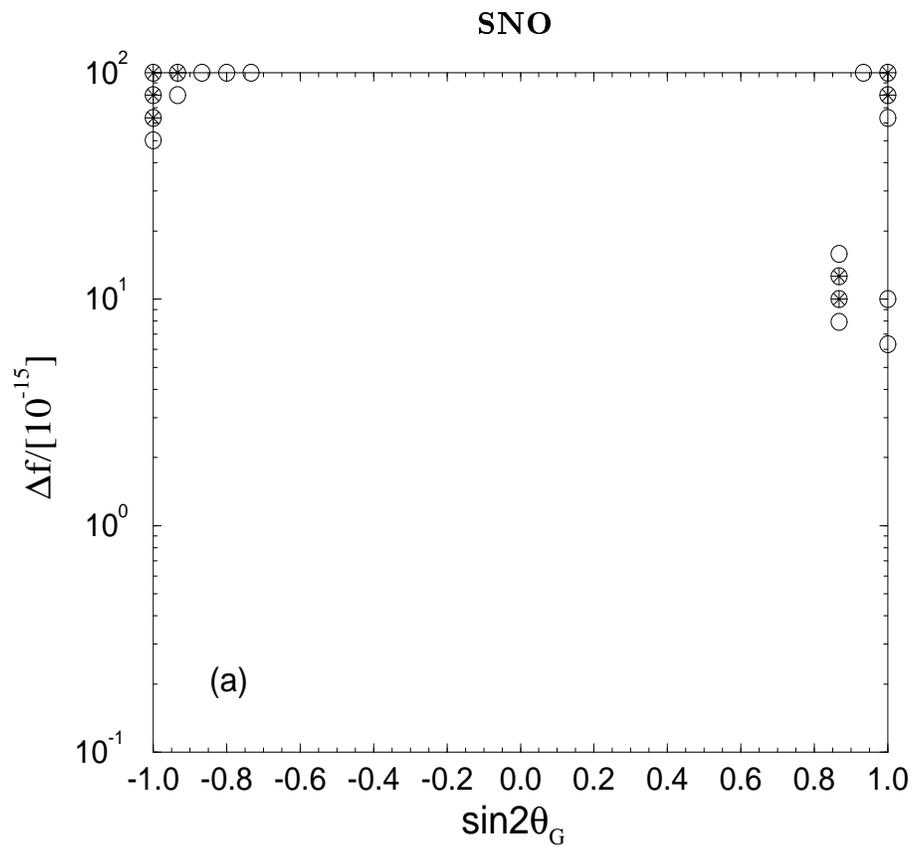
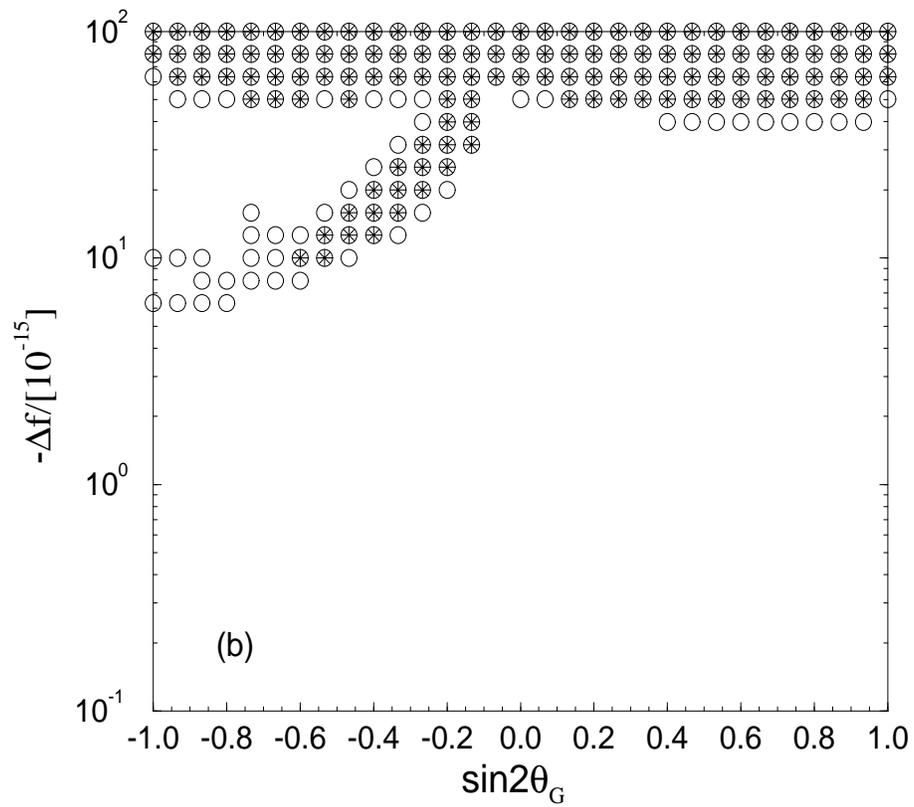
Fig. 6. The same as in Fig. 5 but for SNO. 
The numbers of events are taken as 
12000  (open circles) and  6000  (asterisks). 
\end{document}